\documentclass{article}

\usepackage{arxiv}

\usepackage[utf8]{inputenc} 
\usepackage[T1]{fontenc}    
\usepackage{hyperref}       
\usepackage{url}            
\usepackage{booktabs}       
\usepackage{amsfonts}       
\usepackage{nicefrac}       
\usepackage{microtype}      
\usepackage{lipsum}		
\usepackage{graphicx}
\usepackage{natbib}
\usepackage{doi}
\usepackage{float}
\usepackage{amssymb}
\usepackage{amsmath}
\usepackage{subcaption}
\usepackage{placeins}
\usepackage{wrapfig}

\title{Predictability of bursts of a recurrent nova using topological data analysis and machine learning}

\date{December 31, 2025}	
\author{ \href{https://orcid.org/0009-0004-3459-4772}{\includegraphics[scale=0.06]{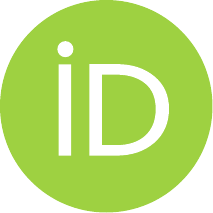}\hspace{1mm}Ignacio ~Morales-Gil}
\\
	Madrid, Spain  \\
	\texttt{ignaciomorgil@gmail.com} \\
}

\renewcommand{\shorttitle}{\textit{arXiv} Template}

\hypersetup{
pdftitle={A template for the arxiv style},
pdfsubject={q-bio.NC, q-bio.QM},
pdfauthor={David S.~Hippocampus, Elias D.~Striatum},
pdfkeywords={First keyword, Second keyword, More},
}

\begin{document}
\maketitle

\begin{abstract}
	RS Oph is a recurrent nova, a kind of cataclismic variable that shows bursts in a period approximately shorter than a century. Persistent homology, a technique from topological data analysis, studies the evolution of topological features of a simplicial complex composed of the data points or an embedding of them, as some distance parameter is varied.
   For this work I trained a supervised learning model based on several featurizations, namely persistence landscapes, Carlsson coordinates, persistent images, and template functions, of the persistence diagrams of sections of the lightcurve of RS Oph.
   A tenfold cross validation of the model based on one of the featurizations, persistence landscapes, consistently shows high recalls and accuracies. This method serves the purpose of predicting whether RS Oph is bursting within a year.
\end{abstract}


\section{Introduction}

Recurrent novae, see \cite{schaefer_review} for a review, are close binary systems, whose primary star, a white dwarf, attracts material from the secondary star, either onto itself or onto an accretion disk. There are two types of mechanism behind the bursts. The first is a thermonuclear runaway explosion. The second is driven by accretion, it may be due to an instability in the accretion disk, or to an unusual mass transfer event, with an increase of the transfer rate or the transfer, of a blob of matter as is the case of RS Oph, see \cite{livio}. If the accreting star is bloated due to having accreted more matter, the accreted material directly impacts the star before turning around it and therefore the lightcurve shows only a primary burst, in contrast to what happens to primary stars that have not yet accreted so much material, like in T Cor B, where the accreting material impinges on an accreting disk or shocks itself after passing around the primary star.

In a burst the magnitude decreases very rapidly, and then swiftly decays back into quiescence, a more stable phase with lower variability, until the next burst.
The prediction of bursts in recurrent novae is currently done comparing some aspects of the lightcurve during the last bursts, like a pre-eruption dip in the case of T Cor B, see \cite{schaefer}. Recurrent novae are not exactly periodic, and existing prediction methods entail big uncertainties, which difficult the prediction of burst times. This difficulties motivate the search for new prediction methods.

Persistent homology, one of the tools of topological data analysis (TDA), see the review of \cite{chazal}, is suitable for this purpose. It has been used to distinguish chaotic and periodic behaviors in dynamical systems, see \cite{khasawneh}. There are several ways to embed time series like a lightcurve so that they have a non-trivial persistent simplicial homology. This topology is defined on a transformation of the original data. In this case, I used an ordinal partition network, which reduces significantly the computational cost. This network is constructed from the permutation sequence that tracks the relative order of the image of several spaced points, as many as the chosen embedding dimension and at the chosen embedding distance from each other, on the lightcurve. Nodes in the ordinal partition network represent the different permutations seen during the slide of the spaced points, while edges, the seen passes between the permutations. A filtration, a nested collection of simplicial subcomplexes, in this case defined on the ordinal partition network, induces the evolution of homology classes respect to the filtration parameter. In this case, the filtration parameter is the diffusion distance, based on random walks through the weighted network, see equation 2.2 in \cite{myersweighted}. These distances then serve to compute the distance matrix with which the persistence diagram is computed. The persistence diagram registers the persistent homology information. It plots the death vs birth times, in terms of the filtration parameter, of homology features of several degrees. Persistence diagrams, in turn, can then be featurized in several ways for machine learning workflows.

Astronomical surveys explore large regions in the sky, some of them repeatedly. This increases their chances of performing the first detection of variable phenomena, such as recurrent novae. When they detect these high energy events, among others, they share an alert with other observatories so they can take more data of the sources. However, research of these phenomena would benefit from predicting these explosions to observe them from the start, without delays or waiting for the jump in magnitude to be first detected. A trustworthy prediction of whether a source, in this work RS Oph, is bursting justifies preparing resources to observe it.

\section{Methods}
The analysis entails loading the data of the lightcurve, choosing the time intervals and assigning them a label according to whether the burst is imminent, still far away or just happened; computing the persistence diagram of the ordinal partition network of the lightcurve in such time intervals, featurizing these persistence diagrams, and performing a cross validation of a supervised machine learning classification, using the featurizations and the labels as ground truth.
\begin{wrapfigure}[32]{r}{0.5\textwidth}
\centering
    \includegraphics[width=0.47\textwidth]{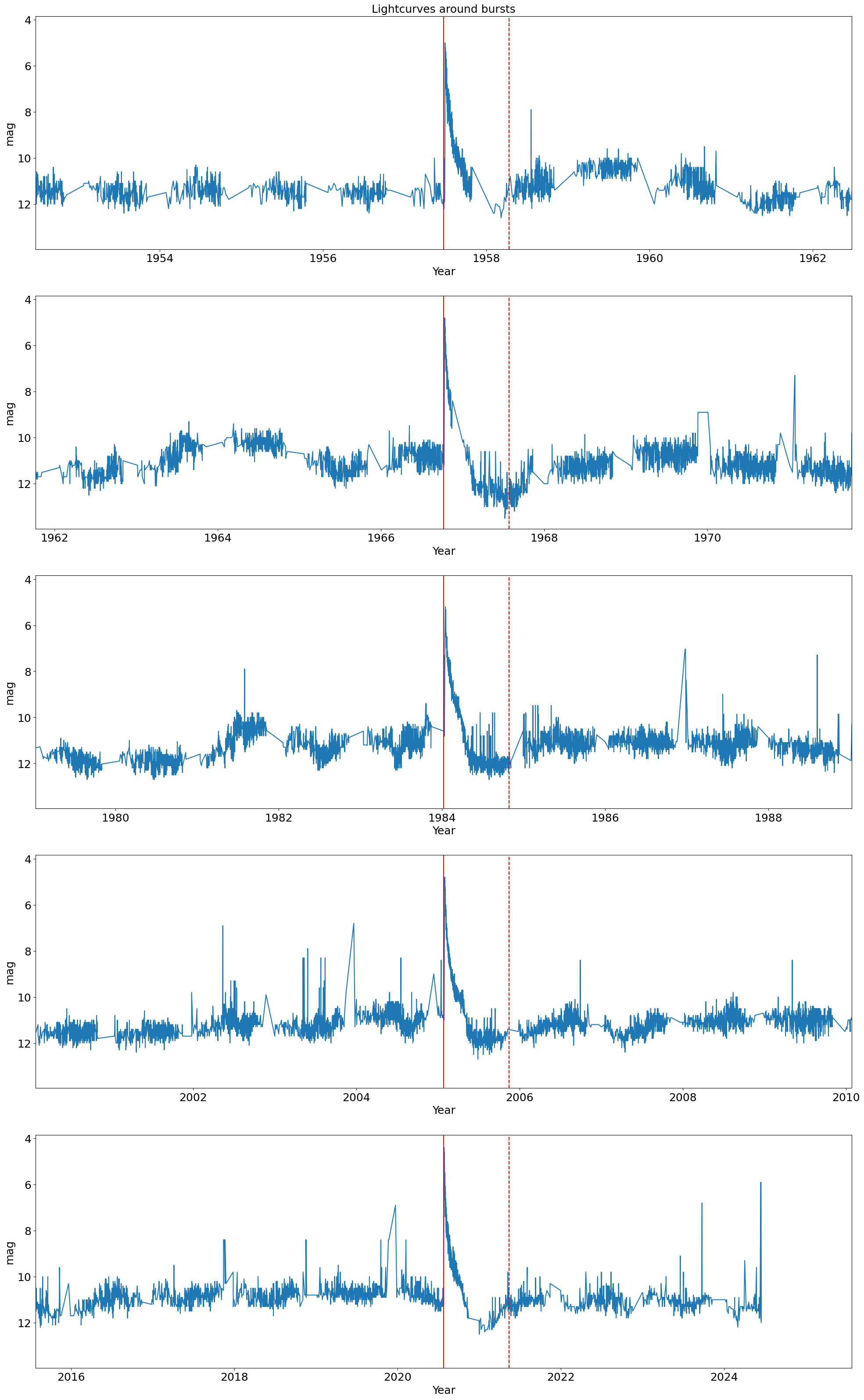}
    \label{bursts}
    \caption{The red solid line is the burst time. The dotted line 0.8 years after the burst marks the approximate end of the decay.}

\end{wrapfigure}

\subsection{Data acquisition and processing}
I downloaded the RS Ophiuchi lightcurve from  AAVSO database, see the webpage in \cite{aavsodownload}. I selected the magnitudes in the band 'Vis', abundant enough in the last decades and encompassing more than five bursts. I also discarded the bound magnitude data (brighter-than magnitudes, $m<m_0$) to avoid big uncertainties, as the effectiveness of the method has been seen to be reduced with higher noise in a similar method, see Results C in \cite{myers}, and there's enough data to describe the lightcurve without brighter-than magnitudes, namely 57266 magnitudes in total. Thus, the magnitude uncertainty, when available (most, 99.8\%, were NAN), is always lower than 0.05 mag.

Then, I manually adjusted the clearest burst times with the guide of the lightcurve, placing the burst times right before the lightcurve begins to peak. The burst times were set, in years, as 1957.480, 1966.770, 1984.020, 2005.070, and 2020.565.

\subsection{Choice of time intervals and labeling}
Then, some time interval ends of the same duration, 4.5 years, are chosen and labeled according to their distance to the five last bursts of the recurrent nova. These last bursts are selected because there is more data of them and there´s no suspicion of having missed a burst between two of them, as \cite{oppenheimer} argues there might have been a missed burst in 1945. 

Concretely, I included 250 linearly spaced points from ten years before and up to seven years after every burst in the list of ends; except for the first of the selected bursts, for which I chose the interval of points from five years before to five years after the burst, to avoid the contamination of the possibly missed burst of 1945. That makes a total of 1950 ends, comprised of 2637 'inter', 378 'pre' and 1985 'post'. Instead of compensating the poorest class with more examples, I used class weighting to avoid overfitting, including the native sklearn option class$\_$weight='balanced' in the models.

Then I labeled the time intervals. If their end is previous to the closest burst, and closer than a year to it, or exactly the burst time, because the intervals are open, I labeled the interval as previous ('pre'); if the end is posterior to the closest burst, and closer to it than an interval duration plus an extra of 0.8 years, that is if the interval contains the burst or its decay, I labeled the interval as posterior ('post'); otherwise I labeled the interval as inter-burst ('inter'). These labels will serve as ground truth for the cross validation. 

\begin{figure}[h]
\includegraphics[width=\hsize]{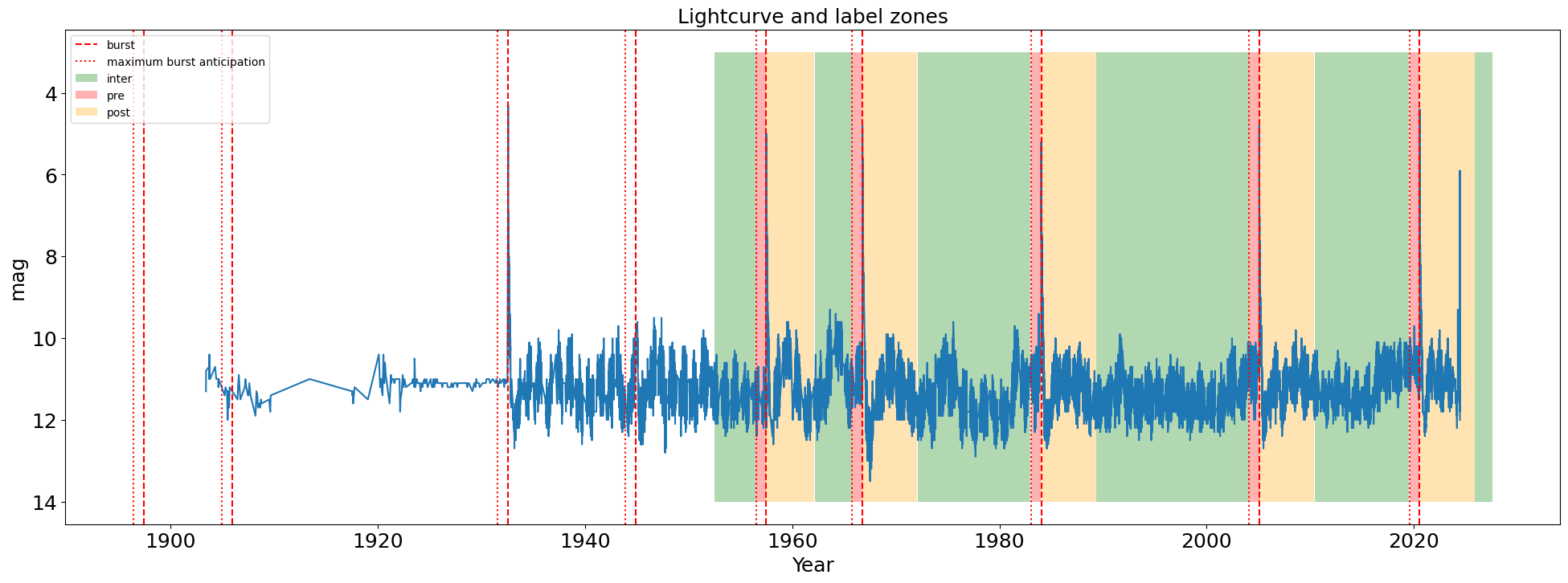}
\label{labelsonlightcurve}
\caption{Dashed vertical lines in red mark the bursts, pointed vertical lines mark the advance time. Areas shaded in red, green and orange mean intervals ending in these areas are labeled as 'pre', 'inter', and 'post', respectively. Some of the first bursts, not included the studied region, are not easily distinguishable.}
\end{figure}

\subsection{Computation of persistence diagrams}
After the time intervals are selected, the sections of the lightcurve with $t\in(\rm{end_i}-\rm{duration},\rm{end_i})$, with $\rm{end_i}$ being each interval open end, are transformed into ordinal partition networks, with embedding dimension $n=5$ and embedding delay $\tau=3$, whose distance matrices, using the diffusion distance, are used to compute their persistence diagrams.

To illustrate the method, I show here three randomly chosen lightcurve sections, belonging to the three classes, 'pre', 'inter' and 'post' respectively; together with their corresponding ordinal partition networks and persistence diagrams. 

\begin{figure}[h!]
     \centering
     \begin{subfigure}[b]{0.3\textwidth}
         \centering
    \includegraphics[width=\textwidth]{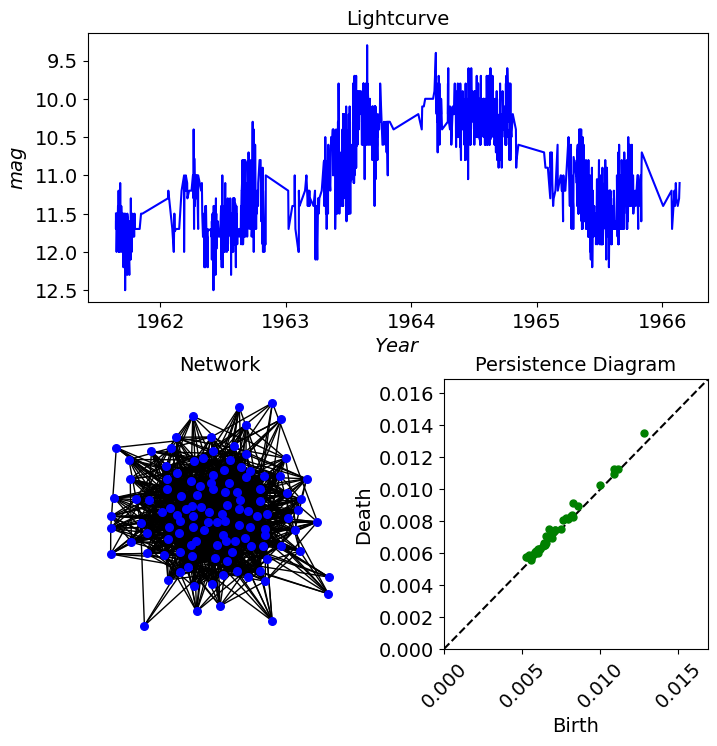}
    \caption{Example of 'pre' network and persistence diagram}
         \label{cont1}
     \end{subfigure}
     \hfill
     \begin{subfigure}[b]{0.3\textwidth}
         \centering
         \includegraphics[width=\textwidth]{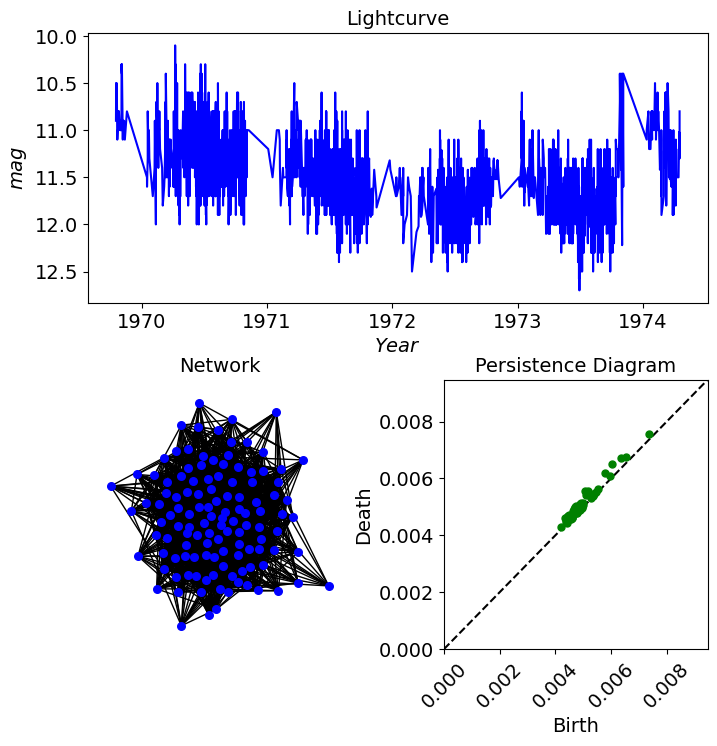}
         \caption{Example of 'inter' network and persistence diagram}
         \label{cont2}
     \end{subfigure}
     \hfill
     \begin{subfigure}[b]{0.3\textwidth}
         \centering
         
    \includegraphics[width=\textwidth]{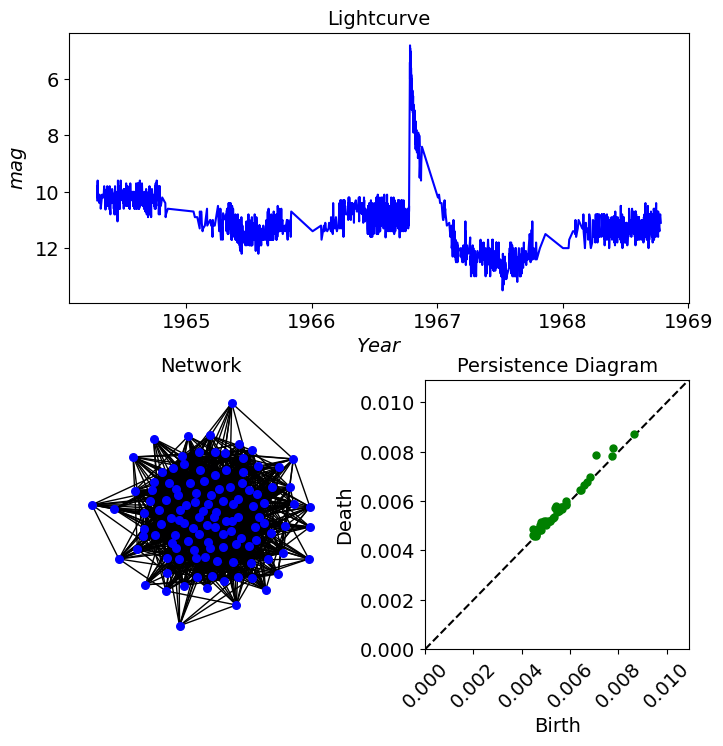}
    \caption{Example of 'post' network and persistence diagram}
    \label{cont3}
    \end{subfigure}
\end{figure}
Only the lightcurve section containing the burst is easily distinguishable from the other classes. Besides, the lightcurve's shape suggests that the energy input that produces the burst has a time scale of a few days, $\Delta t 
\lesssim 3$ days, see section IV of \cite{livio}. Their networks are very hard to distinguish visually as well. Finally, even if one could discern some differences in the persistence diagrams, these diagrams do not have a well defined mean, so the statistical treatment is unavailable. However, a supervised machine learning workflow can classify featurizations of these persistence diagrams.

\subsection{Featurization and cross validation}
 These persistence diagrams are then fed into a tenfold cross validation, see the function getPercentScore at the \cite{tsp_doc_getPercentScore}, based on the sklearn package documented at \cite{pedregosa}, and provided by the teaspoon package described at \cite{khasawneh}. All diagrams were valid, that is, their shapes were acceptable for the function getPercentScore,  so the distribution of labels did not vary, as otherwise some labels would have been discarded with their corresponding diagrams.

For this purpose, getPercentScore internally transforms the diagrams into a featurization that the supervised machine learning workflow can process. There are several such representations. In this work, I used persistence landscapes, see \cite{bubenik}, up to the third function, that is with $\rm{params.PL\_Number}=[1,2,3]$, persistence images, proposed in \cite{adams}, Carlsson coordinates, see \cite{adcock}, with the maximum number of features, that is with $\rm{params.FN}=5$, and finally template functions, as in \cite{perea}, all implemented in the teaspoon package, cf. its documentation at \cite{featurization}. 

I also used a hyperparameter tuning that resulted in much better scores. Particularly, I explored a parameter space for rbf and sigmoid kernels within a logspace from $-3$ to $3$ with 10 values.

Some of these featurizations do have a well-defined mean, like persistence landscapes, which might enable an alternative statistical analysis, though the machine learning approach followed here already obtained very good results.

\section{Results and discussion}

The best results in average accuracy in the test set, as well as in the training set, are obtained from the persistence landscape and template functions featurization  as the table \ref{accuracies_standarddeviations} shows. 
\begin{table}[h!]
\caption{Average accuracies and standard deviations}                 
\label{accuracies_standarddeviations}    
\centering                        
\begin{tabular}{c|c|c}
Featurization	&Test set &Training set 	\\
\hline\hline

Persistence landscapes	&0.824$\pm$ 0.029 &0.874$\pm$ 0.006 	 \\
Template functions	&0.799$\pm$0.042	&0.816$\pm$0.012	 \\
Carlsson coordinates	&0.620$\pm$0.033	&0.627$\pm$0.017	 \\
Persistence images	&0.562$\pm$0.048	&0.563$\pm$0.026	 \\

\end{tabular}
\end{table}

The function getPercentScore with path signature featurization, described in \cite{chevyrev}, and with the kernel featurization, explained in \cite{reininghaus}, was too slow, so I interrupted it before any run had given results. 

The most important metrics for our purpose, evaluating the predictability of bursts of RS Oph, are recall and precision, whose formulae, with "tp" standing for true positives, "fp" for false positives, and "fn" for false negatives, are

\begin{equation}
    \label{eq3}
      \rm{recall}=\frac{\rm{tp}}{\rm{tp}+\rm{fn}} ;~ 
      \rm{precision}=\frac{\rm{tp}}{\rm{tp}+\rm{fp}};
\end{equation}

They measure respectively the fraction of detected positives, and the fraction of rightly diagnosed positives. 

The persistence landscape featurization achieves a precision and recall significantly higher than 50\%. Particularly, without the hyperparameter optimization, the recall result was still compatible with 50\%. The table \ref{precision-recall_standarddeviations_hyperparameterstuned} contains the average results for every featurization, and the results of every run are listed in the appendix.

\begin{table}[h!]
\caption{Average and standard deviation of precision and recall in the test set for the pre-burst class}                 
\label{precision-recall_standarddeviations_hyperparameterstuned}    
\centering                        
\begin{tabular}{c|c|c}
Featurization	&Precision &Recall	\\
\hline\hline

Persistence landscapes	&0.748$\pm$ 0.207 &0.753$\pm$ 0.216 	 \\
Template functions	&0.518$\pm$0.197	&0.647$\pm$0.192	 \\
Carlsson coordinates	&0.239$\pm$0.171	&0.451$\pm$0.289	 \\
Persistence images	&0.081$\pm$0.064	&0.204$\pm$0.163	 \\

\end{tabular}
\end{table}
\FloatBarrier
Thus, the surveillance of the lightcurve would enable the assessment of the source as a target of opportunity in advance, and the preparation of resources to observe the target. The algorithm may be adapted to perform regression instead of classification, predicting the time remaining to the next burst.

This method can be optimized for particular purposes with some variations. The selection of the advance time and duration of the intervals, besides the embedding parameters of the ordinal partition network, or alternatively, using the Takens representation and k-nearest neighbor graph, and the optimization of the featurization parameters may improve the results.


\section{Conclusions}
 Topological data analysis provides featurizations of the lightcurve's persistence diagrams that allow machine learning algorithms to classify the lightcurve sections, effectively predicting whether RSOph is about to burst within a previously established amount of time. 

This method is interesting for surveys and transient alert networks looking for these high energy astronomical bursts, allowing them to point telescopes towards promising targets before the explosion, and registering the event as it explodes, not after it has already begun. This would increase the information about these high energy astrophysical phenomena.

Other applications of TDA in astronomy could be classifying spectra based on these topological measures, complementing the traditional spectral line quotients; superlevel, sublevel or interlevel persistence for the analysis of images; and chromatic persistent homology to study in cosmological simulations or astrometry missions the interaction of several kinds of objects belonging to cosmological structures.
\\
\section{Acknowledgements}
      I acknowledge with thanks the variable star observations from the AAVSO International Database contributed by observers worldwide and used in this research.\\
      This article uses the LaTeX template available at https://github.com/kourgeorge/arxiv-style.\\
      I also thank useful comments from Daniel Nieto Castaño (UCM).

\begin{appendix}

\onecolumn
\section{Appendix. Point summaries}
Point summaries, are scalars computed from their respective persistence diagrams. In this work, besides the supervised learning approach, I obtained for every persistence diagram, the following point summaries: maximum persistence ratio, persistent entropy normalized and homology class ratio from \cite{myers}, besides measures of order like the variance of $H_0$ lengths, the sum of $H_1$ lengths and a linear combination of the last two summaries, the combined persistent homology (CPH$=2\cdot \rm{var}(H_0)$+$\frac{\sum H_1}{2-\sqrt{2}}$) measure, from \cite{motta}, and a close variation, the variance of $H_1$ lengths. However, they don't seem to have a different behavior around bursts, as their time series show.
\begin{figure}[h!]
     \centering
     \begin{subfigure}[b]{0.8\textwidth}
         \centering
\includegraphics[width=\textwidth]
{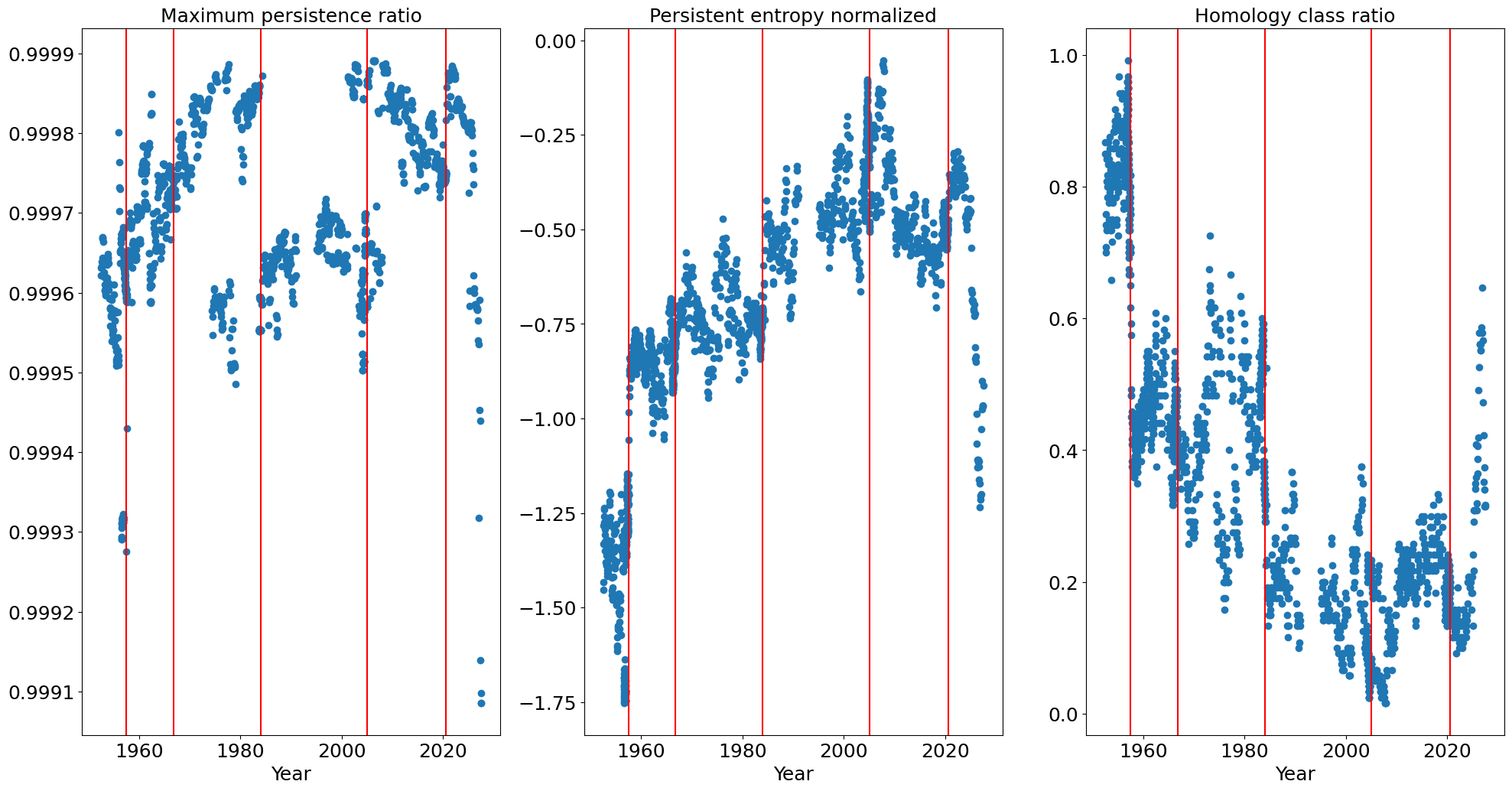}
\label{estadisticospuntuales}
     \end{subfigure}
     \hfill
     \begin{subfigure}[b]{0.8\textwidth}
         \centering
\includegraphics[width=\textwidth]{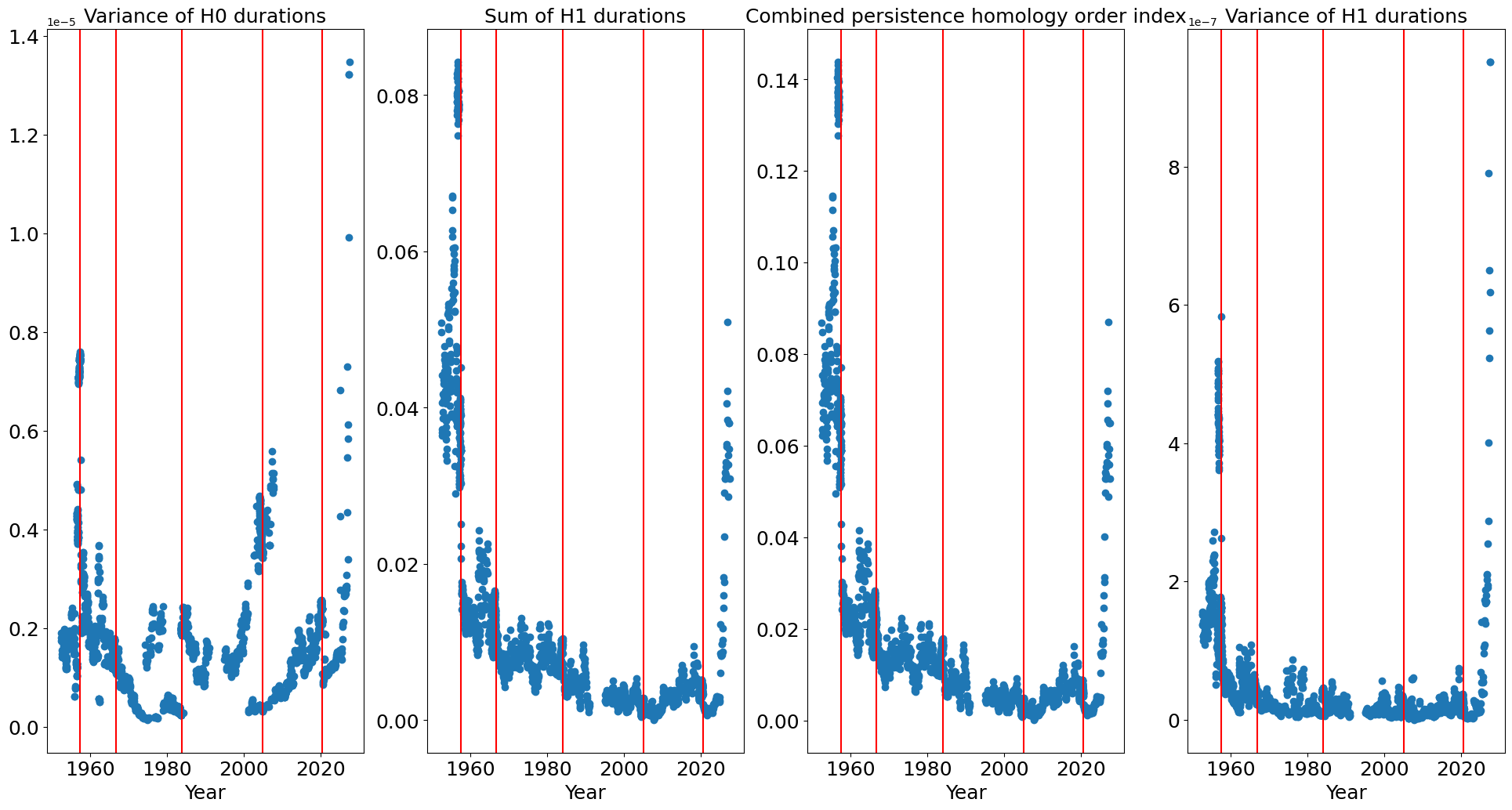}
\label{estadisticosordenhistograma}
\caption{The histograms of CPH and of the sum of $H_1$ durations are very similar because the order of magnitude of the other component in CPH, the variance of $H_0$ durations, is much smaller.}
     \end{subfigure}
     
\end{figure}

In fact their distributions segregated by labels overlap, see the next figures. That is, for most of point summary values, all labels would be similarly likely.

\begin{figure}[h!]
     \centering
     \begin{subfigure}[b]{0.8\textwidth}
         \centering
\includegraphics[width=\textwidth]
{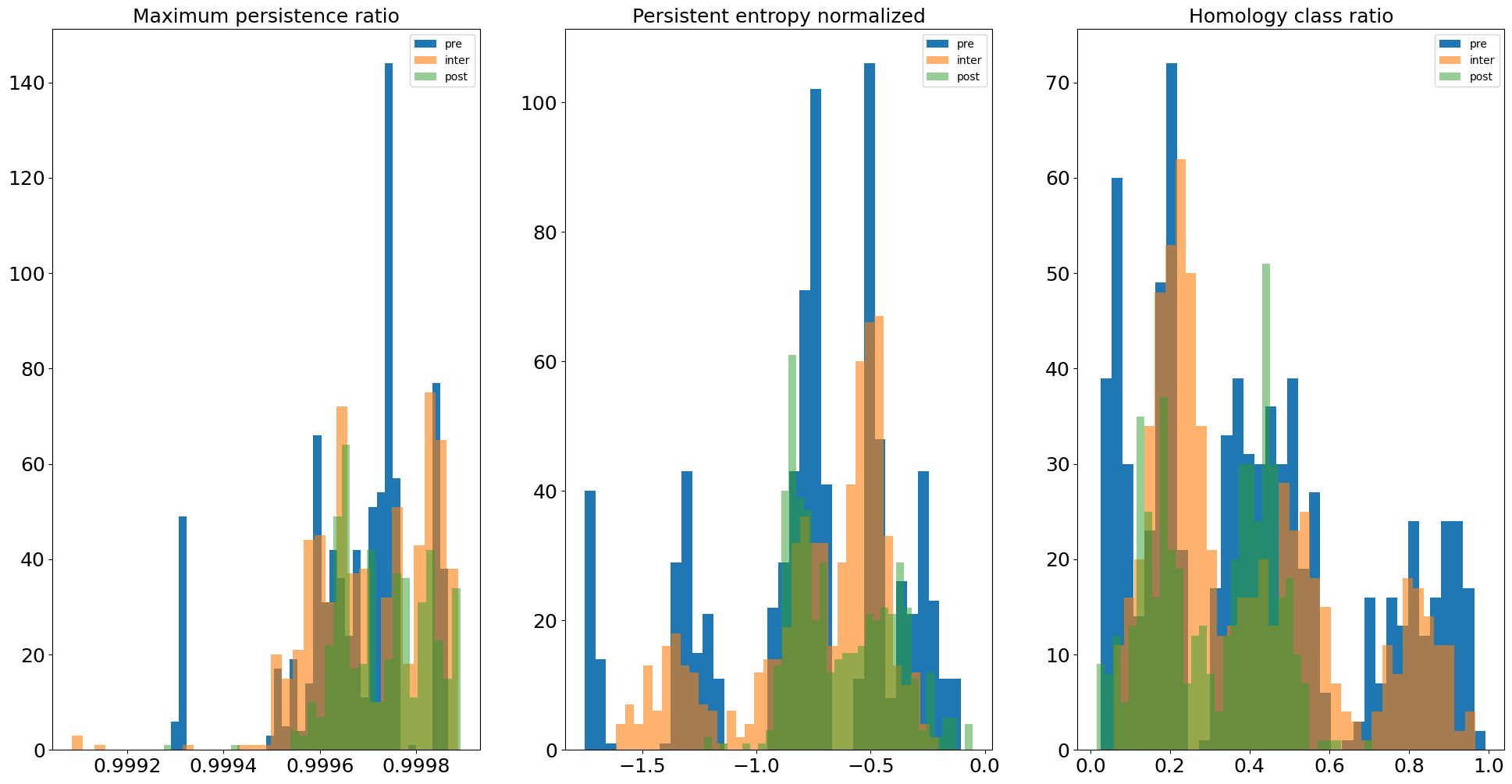}
\label{estadisticospuntualeshistograma}
     \end{subfigure}
     \hfill
     \begin{subfigure}[b]{0.8\textwidth}
         \centering
\includegraphics[width=\textwidth]{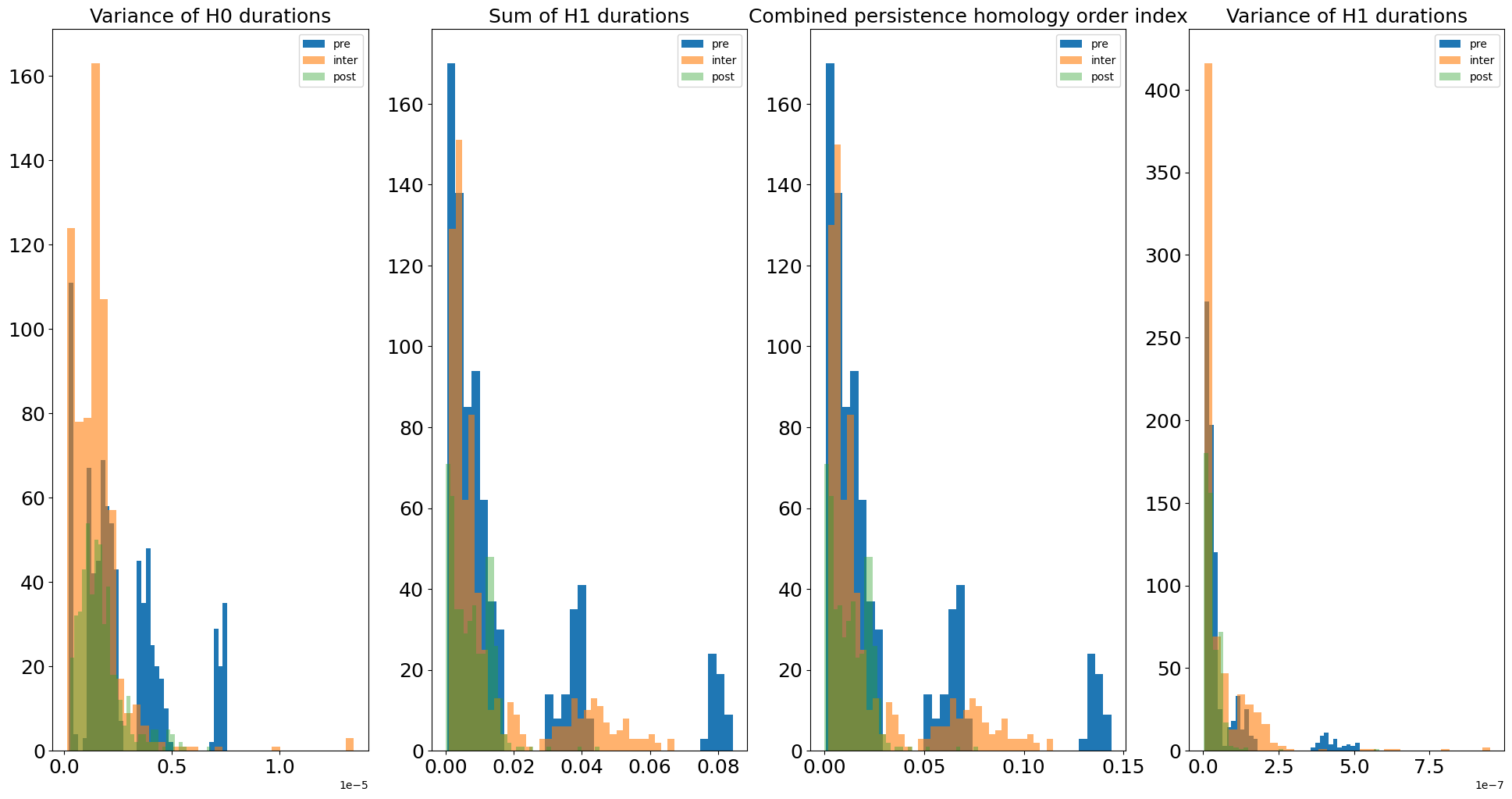}
\label{distribucionesordenhistograma}
\caption{The histograms of CPH and of the sum of $H_1$ durations are very similar because the order of magnitude of the other component in CPH, the variance of $H_0$ durations, is much smaller.}
     \end{subfigure}
     
\end{figure}

\end{appendix}

\begin{appendix}

\onecolumn
\section{Test set precision and recall results for the pre-burst class with hyperparameter tuning}

\begin{table}[h!]
\caption{Accuracies and recalls for the pre-burst cases, with persistence landscape featurization}                 
\label{recalls_accuracies_persistencelandscapes}    
\centering                        
\begin{tabular}{c|c|c|c|c|c|c|c|c|c|c}

Run&1&2&3&4&5&6&7&8&9&10\\
\hline\hline
Precision&0.75&0.667&1&0.5&0.4&1&1&0.667&0.75&1\\
\hline
Recall&0.857&0.5&1&0.4&0.5&0.75&0.857&0.667&1&1\\

\end{tabular}
\end{table} 

\begin{table}[h!]
\caption{Accuracies and recalls for the pre-burst cases, with template functions featurization}                 
\label{recalls_accuracies_templatefunctions}    
\centering                        
\begin{tabular}{c|c|c|c|c|c|c|c|c|c|c}

Run&1&2&3&4&5&6&7&8&9&10\\
\hline\hline
Precision&0.333&0.333&0.429&0.5&0.6&0.714&1&0.4&0.375&0.5\\
\hline
Recall&0.4&1&0.75&0.5&0.857&0.833&0.6&0.5&0.429&0.6\\

\end{tabular}
\end{table} 

\begin{table}[h!]
\caption{Accuracies and recalls for the pre-burst cases, with Carlsson coordinates featurization}                 
\label{recalls_accuracies_templatefunctions}    
\centering                        
\begin{tabular}{c|c|c|c|c|c|c|c|c|c|c}

Run&1&2&3&4&5&6&7&8&9&10\\
\hline\hline
Precision&0.625&0.273&0&0.111&0.286&0.333&0.263&0&0.25&0.25\\
\hline
Recall&0.455&0.75&0&0.333&0.571&0.5&1&0&0.4&0.5\\

\end{tabular}
\end{table}

\begin{table}[h!]
\caption{Accuracies and recalls for the pre-burst cases, with persistent images featurization}                 
\label{recalls_accuracies_templatefunctions}    
\centering                        
\begin{tabular}{c|c|c|c|c|c|c|c|c|c|c}

Run&1&2&3&4&5&6&7&8&9&10\\
\hline\hline
Precision&0.2&0&0.063&0&0.143&0.1&0.125&0&0.067&0.111\\
\hline
Recall&0.286&0&0.2&0&0.5&0.4&0.167&0&0.2&0.286\\

\end{tabular}
\end{table}

\end{appendix}
\begin{appendix}

\onecolumn
\section{Test set precision and recall results for the pre-burst class without hyperparameter tuning}

\begin{table}[h!]
\caption{Average and standard deviation of precision and recall in the test set for the pre-burst class}                 
\label{precision-recall_standarddeviations}    
\centering                        
\begin{tabular}{c|c|c}
Featurization	&Precision &Recall	\\
\hline\hline

Persistence landscapes	&0.80$\pm$ 0.26 &0.65$\pm$ 0.25 	 \\
Template functions	&0.56$\pm$0.22	&0.63$\pm$0.24	 \\
Carlsson coordinates	&0.27$\pm$0.14	&0.63$\pm$0.23	 \\
Persistence images	&0.18$\pm$0.29	&0.25$\pm$0.18	 \\

\end{tabular}
\end{table}

\begin{table}[h!]
\caption{Accuracies and recalls for the pre-burst cases, with persistence landscape featurization}                 
\label{recalls_accuracies_persistencelandscapes}    
\centering                        
\begin{tabular}{c|c|c|c|c|c|c|c|c|c|c}

Run&1&2&3&4&5&6&7&8&9&10\\
\hline\hline
Precision&0.667&0.333&1&1&1&1&0.5&1&1&0.5\\
\hline
Recall&0.8&0.333&0.2&0.833&0.8&0.667&0.667&1&0.4&0.75\\

\end{tabular}
\end{table} 

\begin{table}[h!]
\caption{Accuracies and recalls for the pre-burst cases, with template functions featurization}                 
\label{recalls_accuracies_templatefunctions}    
\centering                        
\begin{tabular}{c|c|c|c|c|c|c|c|c|c|c}

Run&1&2&3&4&5&6&7&8&9&10\\
\hline\hline
Precision&1&0.8&0.429&0.429&0.667&0.6&0.5&0.167&0.625&0.333\\
\hline
Recall&0.5&0.8&0.75&1&0.667&0.75&0.5&0.25&0.833&0.25\\

\end{tabular}
\end{table} 

\begin{table}[h!]
\caption{Accuracies and recalls for the pre-burst cases, with Carlsson coordinates featurization}                 
\label{recalls_accuracies_templatefunctions}    
\centering                        
\begin{tabular}{c|c|c|c|c|c|c|c|c|c|c}

Run&1&2&3&4&5&6&7&8&9&10\\
\hline\hline
Precision&0.25&0.2&0.222&0.429&0.167&0.067&0.368&0.5&0.385&0.111\\
\hline
Recall&0.667&0.4&0.4&0.375&1&0.5&0.778&0.714&1&0.5\\

\end{tabular}
\end{table}

\begin{table}[h!]
\caption{Accuracies and recalls for the pre-burst cases, with persistent images featurization}                 
\label{recalls_accuracies_templatefunctions}    
\centering                        
\begin{tabular}{c|c|c|c|c|c|c|c|c|c|c}

Run&1&2&3&4&5&6&7&8&9&10\\
\hline\hline
Precision&0&0&0&0.25&0.111&0.211&0.091&0.077&1&0.05\\
\hline
Recall&0&0&0&0.3&0.5&0.5&0.333&0.25&0.29&0.333\\

\end{tabular}
\end{table}

\end{appendix}


\begin{thebibliography}{}

    \bibitem[Schaefer(2010)]{schaefer_review} Comprehensive photometric histories of all known galactic recurrent novae, Astrophysical Journal, Supplement Series, ISSN: 00670049, Volume: 187, Issue: 2, Pages: 275 - 373, Article, Open Access, 2010, EID: 2-s2.0-77950592885, DOI: 10.1088/0067-0049/187/2/275, Schaefer B.E.
    
    \bibitem[Livio(1986)]{livio} Livio, M., J. W. Truran, and R. F. Webbink. “A Model for the Outbursts of the Recurrent Nova RS Ophiuchi.” The Astrophysical Journal 308 (September 1986): 736. https://doi.org/10.1086/164546.

    
    \bibitem[Schaefer(2023)]{schaefer}Bradley E Schaefer, The B \& V light curves for recurrent nova T CrB from 1842–2022, the unique pre- and post-eruption high-states, the complex period changes, and the upcoming eruption in 2025.5 ± 1.3, Monthly Notices of the Royal Astronomical Society, Volume 524, Issue 2, September 2023, Pages 3146–3165, https://doi.org/10.1093/mnras/stad735
    
   \bibitem[Chazal(2021)]{chazal}Chazal F and Michel B (2021) An Introduction to Topological Data Analysis: Fundamental and Practical Aspects for Data Scientists. Front. Artif. Intell. 4:667963. doi: 10.3389/frai.2021.667963
   
   \bibitem[Khasawneh(2025)]{khasawneh} Khasawneh, Firas A., Munch, Elizabeth, Barnes, Danielle, Chumley, Max M., Güzel, İsmail, Myers, Audun D., Tanweer, Sunia, Tymochko, Sarah, and Yesilli, Melih. Teaspoon: A Python Package for Topological Signal Processing. Journal of Open Source Software, Vol. 10, No. 107, The Open Journal, p. 7243, March 2025. doi: 10.21105/joss.07243
   
   \bibitem[Myers(2022)]{myersweighted}Myers, Audun, Firas A. Khasawneh, and Elizabeth Munch. “Topological Signal Processing Using the Weighted Ordinal Partition Network.” arXiv:2205.08349. Version 1. Preprint, arXiv, April 27, 2022. https://doi.org/10.48550/arXiv.2205.08349.

   \bibitem[AAVSO(2025)]{aavsodownload}AAVSO webpage. Download Data | aavso, https://www.aavso.org/data-download
   
   \bibitem[Myers(2019)]{myers} Persistent homology of complex networks for dynamic state detection, Audun Myers, Elizabeth Munch, Firas A. Khasawneh, Phys. Rev. E 100, 022314 – Published 21 August, 2019, DOI: https://doi.org/10.1103/PhysRevE.100.022314
   \bibitem[Oppenheimer(1993)]{oppenheimer} Analysis of Long-Term AAVSO Observations of RS Ophiuchi, Volume 22 number 2 (1993), Benjamin D. Oppenheimer, Janet A. Mattei

    \bibitem[teaspoon documentation(2024)]{tsp_doc_getPercentScore} «2.5.3. Classification — teaspoon 1.3.7 documentation». Accessed 9 October, 2025. https://teaspoontda.github.io/teaspoon/modules/ML/CL.html\#teaspoon.ML.P   D\_Classification.getPercentScore.
    
   \bibitem[Pedregosa(2011)]{pedregosa} Scikit-learn: Machine Learning in Python, Pedregosa et al., JMLR 12, pp. 2825-2830, 2011.
   
    \bibitem[Bubenik(2017)]{bubenik} Bubenik, Peter, and Paweł Dłotko. “A Persistence Landscapes Toolbox for Topological Statistics.” Journal of Symbolic Computation, Algorithms and Software for Computational Topology, vol. 78 (January 2017): 91–114. https://doi.org/10.1016/j.jsc.2016.03.009.


    \bibitem[Adams (2017)]{adams} Adams, Henry, Tegan Emerson, Michael Kirby, et al. «Persistence Images: A Stable Vector Representation of Persistent Homology». Journal of Machine Learning Research 18, n.º 8 (2017): 1-35.

    \bibitem[Adcock(2016)]{adcock} Adcock, Aaron, Erik Carlsson, and Gunnar Carlsson. “The Ring of Algebraic Functions on Persistence Bar Codes.” Homology, Homotopy and Applications 18, no. 1 (2016): 381–402. https://doi.org/10.4310/HHA.2016.v18.n1.a21.

    \bibitem[Perea(2023)]{perea} Perea, Jose A., Elizabeth Munch, and Firas A. Khasawneh. “Approximating Continuous Functions on Persistence Diagrams Using Template Functions.” Foundations of Computational Mathematics 23, no. 4 (2023): 1215–72. https://doi.org/10.1007/s10208-022-09567-7.

    \bibitem[Featurization(2024)]{featurization}«2.5.2. Featurization — teaspoon 1.3.7 documentation». Accessed 21 October, 2025. https://teaspoontda.github.io/teaspoon/modules/ML/F\_PD.html\#.

    \bibitem[Chevyrev(2020)]{chevyrev} Chevyrev, Ilya, Vidit Nanda, and Harald Oberhauser. “Persistence Paths and Signature Features in Topological Data Analysis.” IEEE Transactions on Pattern Analysis and Machine Intelligence 42, no. 1 (2020): 192–202. https://doi.org/10.1109/TPAMI.2018.2885516.

   \bibitem[Reininghaus(2015)]{reininghaus} Reininghaus, Jan, Stefan Huber, Ulrich Bauer, and Roland Kwitt. “A Stable Multi-Scale Kernel for Topological Machine Learning.” 2015 IEEE Conference on Computer Vision and Pattern Recognition (CVPR), June 2015, 4741–48. https://doi.org/10.1109/CVPR.2015.7299106.

   
   \bibitem[Motta(2018)]{motta}
    Francis C. Motta, Rachel Neville, Patrick D. Shipman, Daniel A. Pearson, R. Mark Bradley,
    Measures of order for nearly hexagonal lattices,
    Physica D: Nonlinear Phenomena,
    Volumes 380–381,
    2018,
    Pages 17-30,
    ISSN 0167-2789,
    https://doi.org/10.1016/j.physd.2018.05.005.
    (https://www.sciencedirect.com/science/article/pii/S0167278917306851)


\end{thebibliography}
\end{document}